\documentclass[aip,pof,amsmath,amssymb,longbibliography,10pt]{revtex4-1}  % preprint

\usepackage[latin1]{inputenc}
\usepackage[english]{babel}
\usepackage{dcolumn}% Align table columns on decimal point
\usepackage{bm}
\usepackage{amsmath,amssymb,stmaryrd, amsthm}
\usepackage{dcolumn}% Align table columns on decimal point
\usepackage{graphicx}
\usepackage{subfigure}
\usepackage{wasysym}
\usepackage{subeqnarray}
\usepackage{verbatim}
\usepackage{units}
\usepackage{pgf}
\usepackage{natbib}
\usepackage{ulem}

\newcommand{\Eq}{Eq.~}
\newcommand{\Eqs}{Eqs.~}
\newcommand{\Fig}{Fig.~}
\newcommand{\Figs}{Figs.~}
\newcommand{\Tab}{Tab.~}

\newcommand{\cs}{{\mbox{\scriptsize cs}}}
\newcommand{\dem}{{\mbox{\scriptsize DEM}}}
\newcommand{\fs}{{\mbox{\scriptsize f}}}

\newcommand{\rcp}{{\mbox{\scriptsize rcp}}}
\newcommand{\tor}{{\mbox{\scriptsize t}}}
\newcommand{\mer}{{\mbox{\scriptsize m}}}
\newcommand{\vb}{\mathbf{v}}
\begin{document}

% title
\title{Plane shear flows of frictionless spheres: Kinetic theory and 3D Soft-Sphere Discrete Element Method simulations}

% authors
% affiliation: 
% 1 Department, University, City, Postal code, Country
% 2 Corporation or Laboratory, Street address, Postal code, City, Country 
% 3 Department, University, City, State. Zip code, USA 

\author{D. Vescovi}
\affiliation{Department of Civil and Environmental Engineering, Politecnico di Milano, Milan, 20133, Italy.}

\author{D. Berzi}
\affiliation{Department of Civil and Environmental Engineering, Politecnico di Milano, Milan, 20133, Italy.}

\author{P. Richard}%
\affiliation{LUNAM Universit\'e,  IFSTTAR,  site de Nantes, GPEM/MAST, route de Bouaye CS 4, 44344 Bouguenais, France}

\author{N. Brodu}%
\affiliation{Department of Physics, Duke University, Durham, 27708, USA}

%....................................................................................
% abstract
%  An article usually includes an abstract, a concise summary of the work
%  covered at length in the main body of the article. It is used for
%  secondary publications and for information retrieval purposes. 

\begin{abstract}

We use existing 3D Discrete Element simulations of simple shear flows of spheres to evaluate the radial distribution function at contact that enables kinetic theory to correctly predict the pressure and the shear stress, for different values of the collisional coefficient of restitution. 
Then, we perform 3D Discrete Element simulations of plane flows of frictionless, inelastic spheres, sheared between walls made bumpy by gluing particles in a regular array, at fixed average volume fraction and distance between the walls. The results of the numerical simulations are used to derive boundary conditions appropriated in the cases of large and small bumpiness. Those boundary conditions are, then, employed to numerically integrate the differential equations of Extended Kinetic Theory, where the breaking of the molecular chaos assumption at volume fraction larger than 0.49 is taken into account in the expression of the dissipation rate.
We show that the Extended Kinetic Theory is in very good agreement with the numerical simulations, even for coefficients of restitution as low as 0.50. When the bumpiness is increased, we observe that some of the flowing particles are stuck in the gaps between the wall spheres. As a consequence, the walls are more dissipative than expected, and the flows resemble simple shear flows, i.e., flows of rather constant volume fraction and granular temperature.
\end{abstract}

%\pacs{83.80.Fg, 47.57.Gc}% PACS, the Physics and Astronomy

%...................................................................
\maketitle

%-----------------------------------------------------------------
%-----------------------------------------------------------------
\section{Introduction}\label{intro}

Granular materials are collections of discrete particles characterized by loss of energy whenever the particles interact. Their mechanical behavior is very complex even in the case of simple flow conditions (i.e., elementary geometries, stationary motions) or when the granular matter is particularly treatable (i.e., dry, no complex shapes of the grains and no polydispersity, etc).
Due to their microscopic, discrete nature and their macroscopic behavior, granular materials are treated in both the frameworks of discontinuum (Discrete Element simulations) and continuum mechanics. Among the latter, kinetic theories of granular gases represent the most fundamental approach.\\ 
Classic kinetic theories have been derived\cite{jen1983,lun1991,gar1999,gol2003} assuming that the energy of the system is dissipated through binary, instantaneous collisions between smooth spheres, and have been proved to succeed at low to moderate solid volume fractions. When the granular material becomes denser, the assumption of chaotic, binary, instantaneous collisions fails;\cite{cam2002,mit2007,kum2009} also, force chains can develop within the medium.\cite{maj2005,dac2005} Several modifications to the classic kinetic theories have been recently proposed, in order to take into account the role of velocity correlation \cite{jen2006,jen2007} and the development of force chains.\cite{joh1987,joh1990,lou2003,ber2011b,ves2013}\\
The steady plane shear flow of granular materials, in absence of gravity and pressure gradient, serves as test case for the theories. 
Numerical simulations of simple shear flows (i.e., characterized by homogeneous shearing obtained by imposing the Lees-Edwards \cite{lee1972} periodic boundary conditions in the shearing direction) of disks or spheres have been performed using Event-Driven molecular dynamics (ED) and the Soft-Sphere Discrete Element Method (SS-DEM).{\cite{dac2005,ji2006,mit2007,chi2013}}
Inhomogeneous shearing can be obtained in numerical simulations \cite{sho2012,kar2000,liu2005,xu2003} and physical experiments \cite{orl2012,mil2013} when the granular material is sheared between two solid parallel planes, one at rest and the other moving at constant velocity.
In this paper, we numerically solve the Extended Kinetic Theory (EKT), in the form proposed by \citet{ber2013}, for wall-bounded, plane shear granular flows of identical, frictionless spheres, and compare the field variables profiles with those obtained by performing 3D SS-DEM simulations. 
The boundaries are made bumpy by gluing spheres, identical to the moving particles, at the walls; the inter-particle collisions are characterized by the coefficient of restitution, $e$, the ratio of the relative velocity between two impending particles after and before a collision.   
We first use existing numerical results on simple shear flows to slightly modify the constitutive relations of EKT, and then, we analyze both the influence of the coefficient of restitution and the bumpiness. Recently, also \citet{chi2013} have proposed corrections to the kinetic theory of \citet{gar1999} on the basis of 3D SS-DEM simulations of simple shear flows of frictionless and frictional spheres. 
The main differences between our work and theirs are:
(i) we focus on flows where the influence of the boundaries cannot be neglected and propose corrections to the boundary conditions originally developed for nearly elastic spheres glued at the walls \cite{ric1988} for two extreme values of the bumpiness;
(ii) we propose a different radial distribution function obtained from a combination of the classic \citet{car1969}'s and \citet{tor1995}'s expressions, which fits also the numerical data of \citet{chi2013};
(iii) we use an expression for the correlation length recently obtained from the analysis of ED simulations of simple shear flows \cite{ber2013} which does not require additional parameters besides the coefficient of restitution;
(iv) unlike \citet{chi2013}, we show that there is no need for correcting the constitutive relation of the shear stress, provided that the coefficient of restitution is lower than 0.95.\\
The paper is organized as follows.
In Sec. \ref{configuration} we introduce the EKT and the boundary conditions.  
Sec. \ref{simulation} is devoted to describe the simulation method. 
In Sec. \ref{results} we derive the definition of a new radial distribution function and compare the results of the SS-DEM simulations with those obtained from the numerical integration of the equations of EKT.
Finally, concluding remarks are summarized in Sec. \ref{conclusions}.

\newpage
%-----------------------------------------------------------------
%-----------------------------------------------------------------
\section{Governing equations}\label{configuration}

\begin{figure}[!h]
\centering
\includegraphics[width=0.6\textwidth]{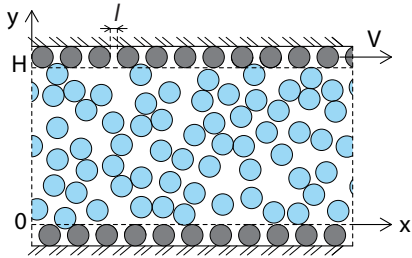}
\caption{Sketch of the constant-volume wall-bounded plane shear flow configuration. A granular material confined between two horizontal solid planes is sheared by moving one plane at constant velocity $V$ ($x$ are $y$ are respectively the flow and shear directions). The two planes are made bumpy by gluing grains at their surface in a regular hexagonal array, where $l$ is the distance between the edges of two adjacent spheres.}
\label{shearflow}
\end{figure}

We focus on the steady motion of a mixture of identical, frictionless spheres sheared between two parallel planes, one at rest and the other moving at constant velocity $V$ (\Fig\ref{shearflow}). \\

%...........................................
%...........................................
\noindent We take $x$ and $y$ to be the flow and the shearing directions, respectively, and ignore variations along the transversal direction $z$. In what follows, all the quantities are made dimensionless using the particle diameter $d$ and density $\rho_p$ and the wall velocity $V$.
Spheres having the same properties of the moving particles are glued at the walls in a regular hexagonal array, where $l$ is the distance between the edges of two adjacent spheres. The bumpiness of the wall is measured by $\psi$, with $\sin\psi = (1+l)/2$.\cite{ric1988} We take $y=0$ to be at the top of the particles glued at the resting wall, and $y=H$ to be at the bottom of the particles glued at the moving wall.

\noindent The hydrodynamic mean fields are the solid volume fraction $\nu$, the velocity along the $x$ direction $u$, the pressure $p$ and the shear stress $s$. For frictionless particles, momentum is exchanged only through collisions,\cite{ves2013} and a description based on kinetic theory \cite{jen1983,lun1991,gol2003} is suitable. We adopt a constant coefficient of restitution $e$.
The continuous velocity field is first coarsed-grained at the scale of a grain diameter. The mean square of the velocity fluctuations against this field, averaged at the same scale, then defines a ``granular temperature'' $T$ field, which measures the degree of agitation of the system.\\
\noindent In the absence of external forces, and in steady conditions, the momentum balance trivially asserts that the pressure and the shear stress are constant along $y$. The balance of the fluctuating energy reads
\begin{equation}\label{energy}
s u' = Q' + \Gamma.
\end{equation}
where $Q$ is the fluctuating energy flux and $\Gamma$ is the rate of dissipation associated to collisions. Here and in what follows, a prime indicates the derivative with respect to the $y$ direction. 
In order to close the problem, we need constitutive relations for $p$, $s$, $Q$ and $\Gamma$. Kinetic theory \cite[e.g.,][]{gar1999} gives 
\begin{equation}\label{pressure}
p = f_1 T,
\end{equation}
\begin{equation}\label{shearstress}
s = f_2 T^{1/2} u',
\end{equation}
\begin{equation}\label{dissipation}
\Gamma = \dfrac{f_3}{L} T^{3/2},
\end{equation}
and
\begin{equation}\label{energyflux}
Q = -f_4 T^{1/2} T' - f_5 T^{3/2} \nu',
\end{equation}
where $f_1, f_2, f_3, f_4$ and $f_5$ are explicit functions of the volume fraction and the coefficient of restitution and are listed in \Tab\ref{tab1}. 
There, $g_0$ is the radial distribution function, whose expression is given in section \ref{results} on the basis of numerical results.
In \Eq\eqref{dissipation}, $L$ is the correlation length, which accounts for the decrease in the rate of collisional dissipation due to the correlated motion of particles that occurs at large volume fraction.\cite{mit2007,kum2009} Taking into account this effect, i.e., the breaking of the molecular chaos assumption, is the peculiarity of EKT.\cite{jen2006,jen2007,jen2010,ber2011a,jen2012,ber2013} The expression for $L$ has been suggested by Jenkins \cite{jen2007} on the basis of a simple heuristic argument,
\begin{equation}\label{func_L}
L = \max\left(1,L^*\dfrac{u'}{T^{1/2}}\right),
\end{equation}
where $L^*$ is a function of the volume fraction and the coefficient of restitution. When $L$ is equal to one, the molecular chaos assumption is valid and EKT reduces to classic kinetic theory. \citet{ber2013} has suggested an expression for $L^*$ on the basis of previous results of ED simulations of simple shear flows:
\begin{equation}\label{Lstar}
L^* = \left(\dfrac{f_2}{f_3}\right)^{1/2}\left[ \dfrac{2\left(1-e\right)}{15}\left(g_0 - g_{0,\fs}\right) +1 \right]^{3/2},
\end{equation}
where $g_{0,\fs}$ is the value of $g_0$ at the freezing point, $\nu = 0.49$, i.e., the lowest value of the volume fraction for which a transition to an ordered state is first possible. \cite{tor1995}

%...........................................
\begin{table}
\centering
\caption{\label{tab1}List of auxiliary coefficients in the constitutive relations of kinetic theory.}
\begin{tabular}{rl}
\hline
\hline
$f_1=$ & $4 \nu G F$ \\[12pt]
$f_2=$ & $ \dfrac{8J}{5\pi^{1/2}} \nu G$\\[12pt]
$f_3=$ & $ \dfrac{12}{\pi^{1/2}}\left(1-e^{2}\right)\nu G$\\[12pt]
$f_4=$ & $\dfrac{4 M \nu G}{\pi^{1/2}}$ \\[12pt]
$f_5=$ & $\dfrac{25\pi^{1/2}N}{128\nu}$ \\[8pt]
$G=$ & $ \nu g_0$\\[8pt]
$F =$ & $\dfrac{1+e}{2} + \dfrac{1}{4G}$\\[12pt]
$J =$ & $\dfrac{1+e}{2} + \dfrac{\pi}{32} \dfrac{ \left[5+2(1+e)(3e-1)G\right]\left[5+4(1+e)G\right]}{\left[ 24-6\left(1-e\right)^2-5(1-e^{2})\right]G^2}$\\[12pt]
$M=$ & $\dfrac{1+e}{2} + \dfrac{9\pi}{144\left(1+e\right)G^2}\dfrac{\left[ 5+3G\left(2e-1\right)\left(1+e\right)^2 \right]\left[5+6G\left(1+e\right)\right]}{16-7\left(1-e\right)}$\\[12pt]
$N=$ & $\dfrac{96\nu\left(1-e\right)}{25G\left(1+e\right)}\dfrac{5+6G\left(1+e\right)}{16+3\left(1-e\right)}\times$\\[12pt]
    & $\left\{\dfrac{20\nu H\left[5+3G\left(2e-1\right)\left(1+e\right)^2\right]}{48-21\left(1-e\right)}-e\left(1+e\right)G\left(1+\nu H\right)\right\}$\\[16pt]
$H=$ & $\dfrac{1}{G}\dfrac{dG}{d\nu}$\\[6pt]
\hline
\hline
\end{tabular}
\end{table}
%...........................................

\noindent From the constitutive relations for the shear stress \eqref{shearstress} and the pressure \eqref{pressure}, we obtain the differential equation governing the velocity,
\begin{equation}\label{dudy}
u' = \dfrac{s}{p}\dfrac{f_1}{f_2}T^{1/2}.
\end{equation}
By deriving \Eq\eqref{pressure} and using \Eq\eqref{energyflux}, the differential equation for the volume fraction results
\begin{equation}\label{dnudy}
\nu' = \dfrac{Q}{T^{1/2}}\dfrac{f_1^2}{f_4}\left[p f_{1,\nu}\left( 1-\dfrac{f_5 f_1}{f_4 f_{1,\nu}} \right)\right]^{-1},
\end{equation}
where $f_{1,\nu}$ represents the derivative of $f_1$ with respect to the volume fraction.
Finally, using \Eqs\eqref{energy}, \eqref{pressure}, \eqref{dissipation} and \eqref{dudy}, the differential equation for the energy flux reads
\begin{equation}\label{dQdy}
Q' = p T^{1/2}\left[\dfrac{f_1}{f_2} \left(\dfrac{s}{p} \right)^2 - \dfrac{f_3}{L f_1} \right]. 
\end{equation}
We also introduce an additional differential equation for the partial mass hold-up, defined as $m = \int_0^y \nu dz$, 
\begin{equation}\label{dmdy}
m' = \nu.
\end{equation}
Then, the value of the average volume fraction $\bar\nu$ along $y$ can be implemented as a boundary condition for $m$.
 
\noindent We numerically solve the set of the four differential equations \Eqs\eqref{dudy}-\eqref{dmdy} using the function `bvp4c' implemented in MATLAB, and fixing the gap $H$. We treat the pressure and the shear stress as parameters, so that we need six boundary conditions to solve the problem. As already mentioned, we implement the fixed average volume fraction as a boundary condition for the partial mass hold-up, i.e., $m_H = \bar\nu H$, while, at the resting wall, $m_0=0$. Here and in what follows, the index represents the coordinate $y$ at which the quantity is evaluated. We allow the particles to slip at the bumpy walls, so that, for symmetry, $u_0 = u_w$ and $u_H = 1-u_w$, where \citet{ric1988} obtained, in the case of rigid, nearly elastic semi-spheres attached to a flat wall,
\begin{equation}\label{slipvelocity}
u_w = \sqrt{\dfrac{\pi}{2}}h\dfrac{s}{p}T_0^{1/2},
\end{equation}
with
\begin{equation}\label{func_h}
h = \dfrac{2}{3}\dfrac{\left[ 1+\dfrac{5 F_0 \left( 1+ B\right)\sin^2\psi}{2\sqrt{2}J_0} \right]}{\dfrac{2\left(1-\cos\psi\right)}{\sin^2\psi}}+\dfrac{5 F_0}{\sqrt{2}J_0},
\end{equation}
where $B = \pi\left[1+5/\left(8G_0\right)\right]/\left(12\sqrt{2}\right)$, and $J_0$, $F_0$ and $G_0$ are obtained from the corresponding expressions of \Tab\ref{tab1} with $\nu = \nu_0$. The bumpy walls act either as a sink or a source of fluctuating energy to the system. The two boundary conditions for the energy flux are $Q_0=Q_w$ and $Q_H=-Q_w$, where \citet{ric1988} proposed
\begin{equation}\label{flux_bc}
Q_w = s u_w - \sqrt{\dfrac{\pi}{2}}p T_0^{1/2}(1-e)\dfrac{2\left(1-\cos\psi\right)}{\sin^2\psi}.
\end{equation}
The results of the numerical integration will be compared with those obtained from SS-DEM simulations described in Sec. \ref{simulation}.

%-----------------------------------------------------------------
%-----------------------------------------------------------------
\section{{SS-DEM} Simulations}\label{simulation}

\begin{figure}[!h]
\centering
\includegraphics[width=0.5\textwidth]{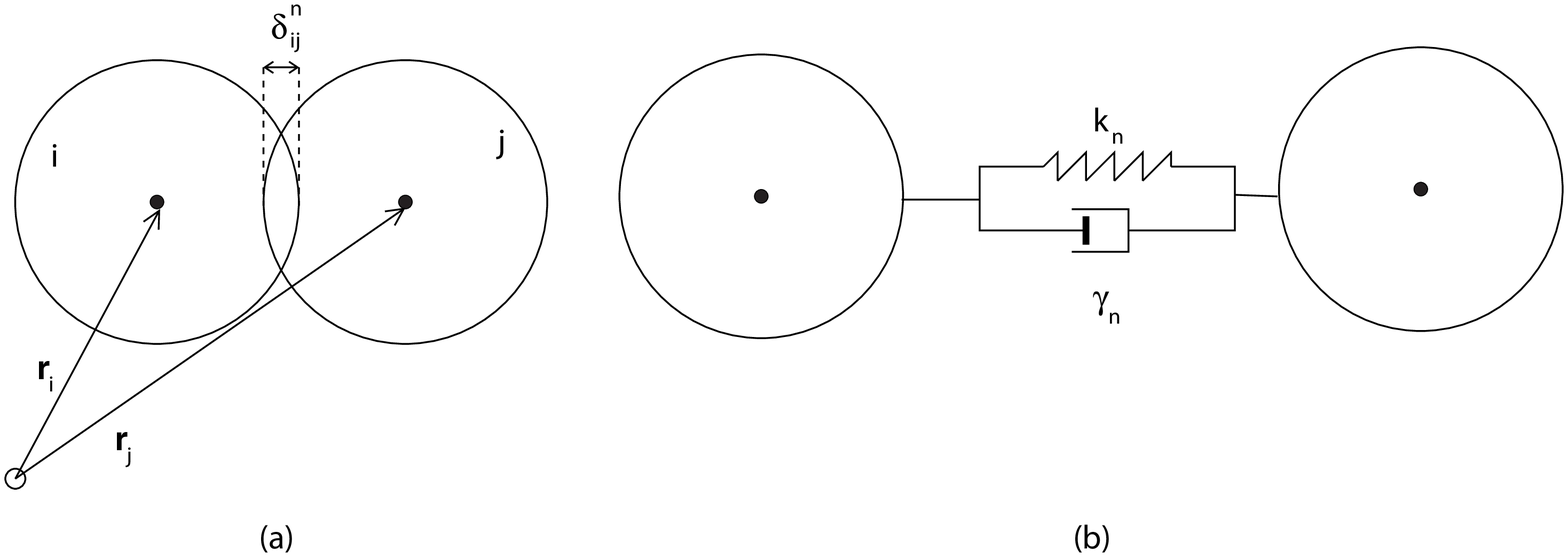}
\caption{Sketches of two particles at contact (a) and of the contact forces used (b)}
\label{forcesfig}
\end{figure}

\begin{figure}[!h]
\centering
\includegraphics[width=0.7\textwidth]{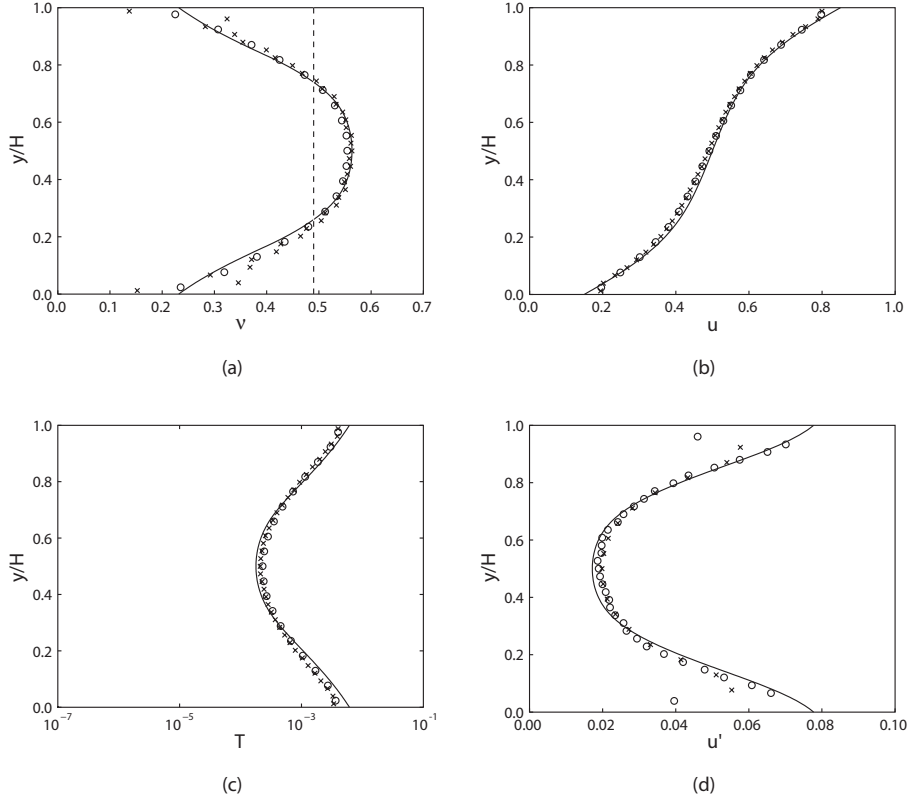}
\caption{Profiles of $\nu$, $u$, $T$ and $u'$ obtained from SS-DEM simulation when $H = 18$, $\bar\nu = 0.45$, $\psi = \pi/5$ and $e = 0.80$, when the domain along the $y$-direction is divided into 20 (open circles) and 40 (crosses) slices to perform the averaging. The solid lines are the results of EKT when \Eq\eqref{g0_theo} and \Eq\eqref{func_L} are employed. The dashed line in (a) is the value of the volume fraction at the freezing point, $\nu=0.49$.}
\label{examples}
\end{figure}

We have carried out 3D SS-DEM simulations using our own code~\cite{Brodu_PRE_2013} to make comparisons with the results of EKT. 
Although this method is well known and can be found in many papers,\cite{sil2001,Taberlet2008,Taberlet2006,Rycroft2009,Majmudar2007,Hirshfeld2001,Richard2012,Richard_PRL_2008} 
we present it here to support our discussion on the comparison of numerical and theoretical results (\Fig\ref{forcesfig}).
%...........................................
%...........................................
In this method, each grain $i$ is a soft sphere  of diameter $d_i$, mass $m_i$, moment of inertia $I_i$, position $\mathbf{r}_i$, velocity $\mathbf{v}_i$ and angular velocity ${\boldsymbol{\omega}}_i$. For a pair of particles
$\left\{i,j\right\}$, we define the relative distance vector $\mathbf{r}_{ij}=\mathbf{r}_i-\mathbf{r}_j$, their separation $r_{ij}=\left|{\mathbf{r}}_{ij}\right|$, the relative velocity ${\mathbf{v}}_{ij}={\mathbf{v}}_{i}-{\mathbf{v}}_j$, and the normal unit vector ${\mathbf{n}}_{ij}=({\mathbf{r}}_i-{\mathbf{r}}_j) / r_{ij}$. These two particles are in contact if their normal overlap $\delta_{ij}^n=\mbox{max}(0,d_i/2+d_j/2-r_{ij})$ is strictly positive. 
%The contact point is then assumed to be the center of the overlap. 
In general, the force on particle $i$ from the interaction with particle $j$ is the sum of a normal and tangential contribution : $\mathbf{f}_{ij}=\mathbf{f}^n_{ij}+\mathbf{f}^t_{ij}.$ 
However, the present work deals with frictionless particles for which the contact force is purely normal. 
Therefore, the grains are submitted to neither tangential forces nor torques. 
For the normal force, we use the standard spring-dashpot interaction model:\cite{Luding2008} 
$\mathbf{f}^n_{ij}=k_n\delta_{ij}^n\mathbf{n}_{ij}-\gamma_n\mathbf{v}_{ij}^n,$
where $k_n$ is the spring constant, $\gamma_n$ the damping coefficient and $\mathbf{v}_{ij}^n$ the normal relative velocity $\mathbf{v}_{ij}^n=(\mathbf{v}_{ij}\cdot\mathbf{n}_{ij})\mathbf{n}_{ij}.$ The damping is used to obtain an inelastic
collision. 
For a purely normal collision, the collision time $t_c$ is equal to 
$\pi/\left[k_n/m_{ij} - \gamma_n^2/\left(4m_{ij}^2\right)\right]^{1/2}$, with the reduced mass $m_{ij}=m_im_i/(m_i+m_j)$. The normal restitution coefficient is given by
$e=\exp\left[-t_c\gamma_n/(2m_{ij})\right].$
The total force on particle $i$ is then a combination of contact forces with other particles and the boundaries and an eventual resulting external force $\mathbf{F}_{ext}$. 
The resulting force $\mathbf{f}_i$ is given by 
\[
\mathbf{f}_i=\mathbf{F}_{ext}+\sum_{j=1,j\neq i}^N\mathbf{f}_{ij},
\]
where $N$ is the total number of flowing spheres.
Once the forces are calculated for all the particles, the Newton's equations of motion, $m_i{d^2\mathbf{r}_i}/{dt^2} = \mathbf{f}_{i}$, for the  translational degrees of freedom are integrated.  
We use a velocity Verlet integration scheme with a time step $\Delta t=t_c/30$. The grains have all the same size and density. As already mentioned, the numerical results are given in nondimensional units: distances, times, velocities, forces, elastic constants and viscoelastic constants are,
respectively, measured in units of $d$, $d/V$, $V$, $\rho_p d^2V^2$,  $\rho_p dV^2$ and $\rho_p d^2 V$.\\
\noindent All the simulations have been performed in a rectangular box of length $L_x = 20$, width $L_z = 10$ and height $L_y=20$ - so that $H = L_y-2 = 18$ - with $N=3132$. The bumpiness has been generated by gluing, in a regular hexagonal array, a total of 340 particles at the two walls in the case of $\psi=\pi/5$, and 154 in the case $\psi=\pi/3$. Hence, taking into account the extra-space accessible to any flow particle in between the wall-spheres, $\bar\nu=0.45$ when $\psi=\pi/5$ and $\bar\nu=0.44$ when $\psi=\pi/3$. The particle stiffness of the linear spring model has been set equal to $2\cdot 10^5$. The non-dimensional ratio of the particle stiffness over the particle pressure is greater than $10^5$ in all the simulations. This ensures that the contact time during a collision is much less than the flight time in between two successive collisions, so that the latter can be considered instantaneous.\cite{ots2010,ber2011b} The value of $\gamma_n$ is adjusted to obtain the chosen normal restitution coefficient. Periodic boundary conditions are employed in the $x$ and $z$ directions and the horizontal flat walls are located at $y=-1$ and $y=H+1$, the latter moving at constant horizontal velocity $V$. 
Those walls are treated as spheres of infinite size and density and the grains glued on their surface to create the bumpiness are treated like spheres of diameter $1$ and infinite density.\\
\noindent We focus on the steady state of sheared granular flows, that we consider achieved when the space-averaged granular temperature $\bar{T}$ becomes approximately constant (fluctuations around the time-averaged value less than $10\%$).
The space-averaged granular temperature is computed as
\[
 \bar{T} = \dfrac{1}{3N}\left[ \displaystyle\sum_{i=1}^N  \left\|\vb_i\right\|^2 - \left( \displaystyle\sum_{i=1}^N  \left\|\vb_i\right\| \right)^2\right],
\]
where $\left\|\cdot\right\|$ denotes the Euclidean norm of a vector. 
Simulations have been performed by changing the coefficient of restitution ($e=0.20$, 0.50, 0.60, 0.70, 0.80, 0.92, 0.98) and the bumpiness ($\psi=\pi/5$ and $\pi/3$).\\
We have checked that the steady state does not depend on the initial configuration, by preparing two different initial states, consisting of $N$ spheres uniformly distributed in the volume. In the first case the spheres are initially at rest; in the second case, we assign a linear distribution (from 0 to 1) of the $x$-velocity of the spheres. This second configuration corresponds to a higher value of the initial energy, i.e., of the initial space-averaged granular temperature. In both cases, we have achieved the same steady state, i.e., with the same value of space-averaged granular temperature and the same distributions of the field variables.
%...........................................
\begin{figure}[!h]
\centering
\includegraphics[width=0.7\textwidth]{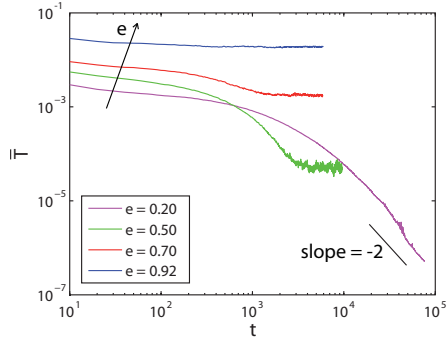}
\caption{Time evolution of the mean granular temperature for different values of the coefficient of restitution when $\bar\nu = 0.45$ and $\psi = \pi/5$.}
\label{evolution_pi5}
\end{figure}
%...........................................
The time at which the steady state is reached increases when the coefficient of restitution decreases (e.g., see \Fig\ref{evolution_pi5} for the case $\psi = \pi/5$). For sufficiently small coefficients of restitution (case $e=0.2$ in \Fig\ref{evolution_pi5}), the mean granular temperature continues to decrease, without reaching a steady state. The slope of the curve approaches the value -2 that characterizes the Homogeneous Cooling State (HCS),\cite{gol2003} where the rate of change of the granular temperature in the balance of fluctuating energy is only due to the collisional dissipation and the granular temperature obeys the Haff's law, \cite{haf1983} $T\propto\left(1+t \right)^{-2}$. We will discuss in Sec. \ref{results} this finding. \\
Once the steady state is reached, measurements are averaged in time, over at least 2000 time steps, and over the lengths of the domain along the $x$ and $z$ directions, using 20 horizontal slices. Given that the averaging is sensitive to the amplitude of the spatial discretization,\cite{wei2013} we chose a number of slices that does not affect the results.
Example of profiles of $\nu$, $u$, $T$ and $u'$ are plotted in \Fig\ref{examples} for $\psi = \pi/5$ and $e = 0.80$, when 20 or 40 horizontal slices are employed. Also shown are the results of the numerical integration of the EKT described in Sec. \ref{results}. 
The velocity profile has a characteristic S-shape, in agreement with recent physical experiments performed on disks.\cite{mil2013} Also, the profile of the shear rate very much resembles the experimental findings. The volume fraction increases and the granular temperature decreases with distance from the walls. The core of the flow is dense, i.e., the volume fraction is larger than 0.49, and there the molecular chaos assumption breaks down. All those features are well captured by kinetic theory. 

%...........................................
%...........................................

%-----------------------------------------------------------------
%-----------------------------------------------------------------
\section{Results and comparisons}\label{results}

\begin{figure}[!h]
\centering
\includegraphics[width=0.7\textwidth]{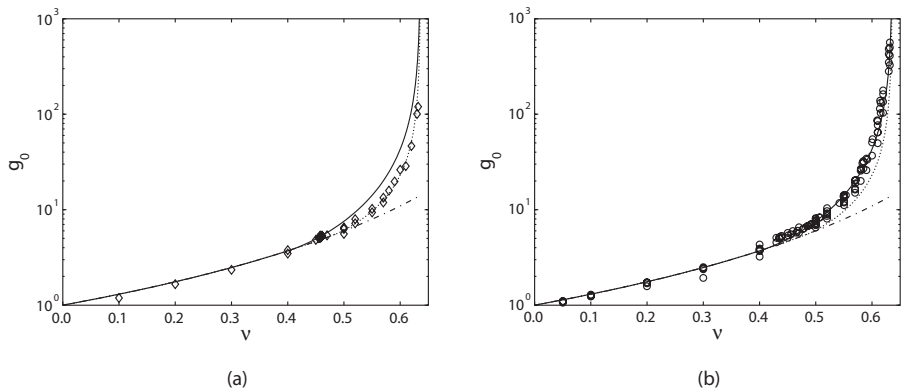}
\caption{Numerical (symbols) radial distribution function (after \citet{mit2007}, \citet{chi2013} and present SS-DEM simulations) as a function of the volume fraction for: (a) $e=0.98$ and 0.99; (b) $0.5\leq e\leq 0.95$. Also shown are \Eq\eqref{g0_theo} (solid line) and the expressions of Carnahan and Starling (dot-dashed line) and Torquato (dotted line).}
\label{fig_g0}
\end{figure}

We first use the numerical results obtained on simple shear flows of frictionless spheres by \citet{mit2007} and \citet{chi2013} to derive the expression of the radial distribution function $g_0$. \citet{mit2007} performed ED simulations of inelastic hard spheres, whereas \citet{chi2013} used a SS-DEM code with a linear spring-dashpot model. In both works, the Lees-Edwards\cite{lee1972} boundary conditions were implemented in the shearing direction, in order to allow for the system to remain homogeneous during the shearing. 
From the constitutive relation for the pressure \eqref{pressure} and the expressions of \Tab\ref{tab1},
\begin{equation}\label{g0}
g_0 = \dfrac{1}{2\nu(1+e)}\left(\dfrac{p}{\nu T}-1\right),
\end{equation}
so that the radial distribution function can be obtained from the numerical values of pressure, volume fraction and granular temperature.
For small volume fractions, $g_0$ obeys the Carnahan and Starling's expression,\cite{car1969}
\begin{equation}\label{carnahanstarling}
g_{0,\cs} = \dfrac{2-\nu}{2\left(1-\nu\right)^3},
\end{equation}
whereas Torquato's\cite{tor1995} proposed, on the basis of numerical results on elastic particles,
\begin{equation}\label{torquato}
g_{0,\tor} = \begin{cases}
\begin{array}{lr}
g_{0,\cs} & \mbox{ if } \nu < 0.49,\\
\dfrac{\left(2-0.49\right)}{2\left(1-0.49\right)^3}\dfrac{\left(\nu_\rcp-0.49\right)}{\left(\nu_\rcp-\nu\right)} &  \mbox{ otherwise.}
\end{array}
\end{cases}
\end{equation}
with $\nu_\rcp = 0.636$ the value of volume fraction at random close packing.
\Fig\ref{fig_g0} shows the radial distribution function obtained from the numerical simulations of \citet{mit2007} and \citet{chi2013} on simple shear flows and the present SS-DEM simulations of bounded shear flows. \Eq\eqref{torquato} fits well the numerical results in the case of nearly elastic particles (\Fig\ref{fig_g0}a), while underestimates the data for dense flows of particles when $e \leq 0.95$ (\Fig\ref{fig_g0}b). In the latter case, we propose to use the following expression:
\begin{equation}\label{g0_theo}
g_0 = f g_{0,\cs} + \left(1 - f\right)\dfrac{2}{\nu_\rcp - \nu},
\end{equation}
where $f$ is a function of the volume fraction which makes $g_0$ equal to the Carnahan and Starling's expression when the volume fraction is less than a limit value, $\nu_\mer$, 
\begin{equation}\label{func_f}
f = \begin{cases}
\begin{array}{lr}
1 & \mbox{ if } \nu < \nu_\mer,\\
\dfrac{\nu^2-2\nu_\mer\nu +\nu_\rcp\left(2\nu_\mer-\nu_\rcp\right)}{2\nu_\rcp\nu_\mer -\nu_\mer^2-\nu_\rcp^2} &  \mbox{ otherwise.}
\end{array}
\end{cases}
\end{equation}
We take $\nu_\mer = 0.4$; the quadratic expression for $f$ when $\nu \geq \nu_\mer$ ensures that the first derivative of $g_0$ is continuous, facilitating the numerical integration of the equations.

%...........................................
%...........................................

\begin{figure}[!h]
\centering
\includegraphics[width=0.7\textwidth]{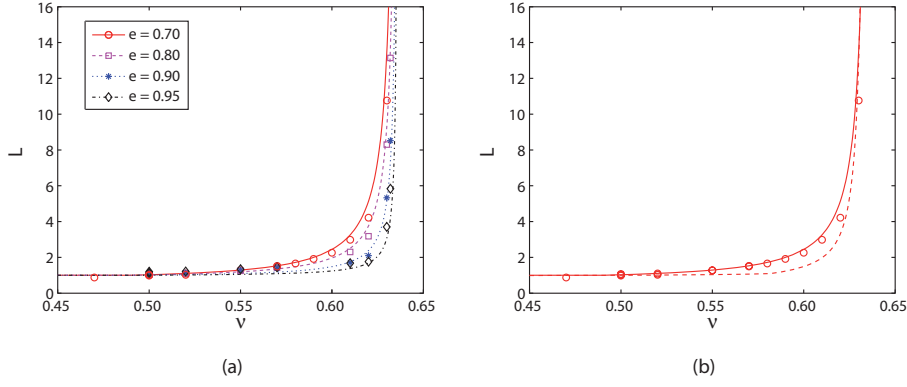}
\caption{(a) Numerical (symbols, after \citet{mit2007} and \citet{chi2013}) and theoretical (lines, \Eq\eqref{ssf_L}) correlation length as a function of the volume fraction, for different values of the coefficient of restitution: $e = 0.70$ (circles and solid line); $e = 0.80$ (squares and dashed line); $e = 0.90$ (stars and dotted line); $e = 0.95$ (diamonds and dot-dashed line). (b) Same as in \Fig\ref{fig_L}(a) for the case $e=0.7$. The dashed line represents the theory of \citet{chi2013}.}
\label{fig_L}
\end{figure}

\noindent In simple shear flows, the divergence of the flux of fluctuating energy in \Eq\eqref{energy} can be neglected, and the correlation length reduces to:
\begin{equation}\label{ssf_L}
L = \dfrac{f_3 T^{3/2}}{s u'}.
\end{equation}
In \Fig\ref{fig_L} we plot the quantity $f_3 T^{3/2}/\left(s u'\right)$ as a function of the volume fraction, where $s$ and $T$ are those measured by \citet{chi2013} in their SS-DEM simulations, while $f_3$ is evaluated from the expression of \Tab\ref{tab1}, using \Eq\eqref{g0_theo} and the measured values of the volume fraction. There, the lines represent the theoretical expression of the correlation length, which, in simple shear flows, using \Eqs\eqref{shearstress},\eqref{func_L} and \eqref{ssf_L}, is
\begin{equation}\label{ssf_L2}
L = \max\left(1,\dfrac{f_2^{1/3}}{f_3^{1/3}} {L^*}^{2/3}\right).
\end{equation}
The agreement between the numerical data and the theoretical expression of $L$ is remarkable. In \Fig\ref{fig_L}(b) we also plot, for comparison, the correlation length obtained from the modification of the kinetic theory suggested by \citet{chi2013} when $e=0.7$.

\noindent In simple shear flows, an algebraic relation between the shear rate and the granular temperature exists,
\begin{equation}\label{ssf_T}
\dfrac{T}{{u'}^2} = \dfrac{f_2}{f_3}L.
\end{equation}
Substituting \Eq\eqref{ssf_T} in \Eq\eqref{pressure} and \Eq\eqref{shearstress} leads to the following expressions for the pressure,
\begin{equation}\label{ssf_pressure}
p = f_1\dfrac{f_2}{f_3} L {u'}^2,
\end{equation}
the shear stress
\begin{equation}\label{ssf_shearstress}
s = \left(\dfrac{f_2^3}{f_3}L\right)^{1/2} {u'}^2,
\end{equation}
and the stress ratio
\begin{equation}\label{ssf_stressratio}
\dfrac{s}{p} = \left(\dfrac{f_2f_3}{f_1^2L}\right)^{1/2}.
\end{equation}

\begin{figure}[!h]
\centering
\includegraphics[width=0.7\textwidth]{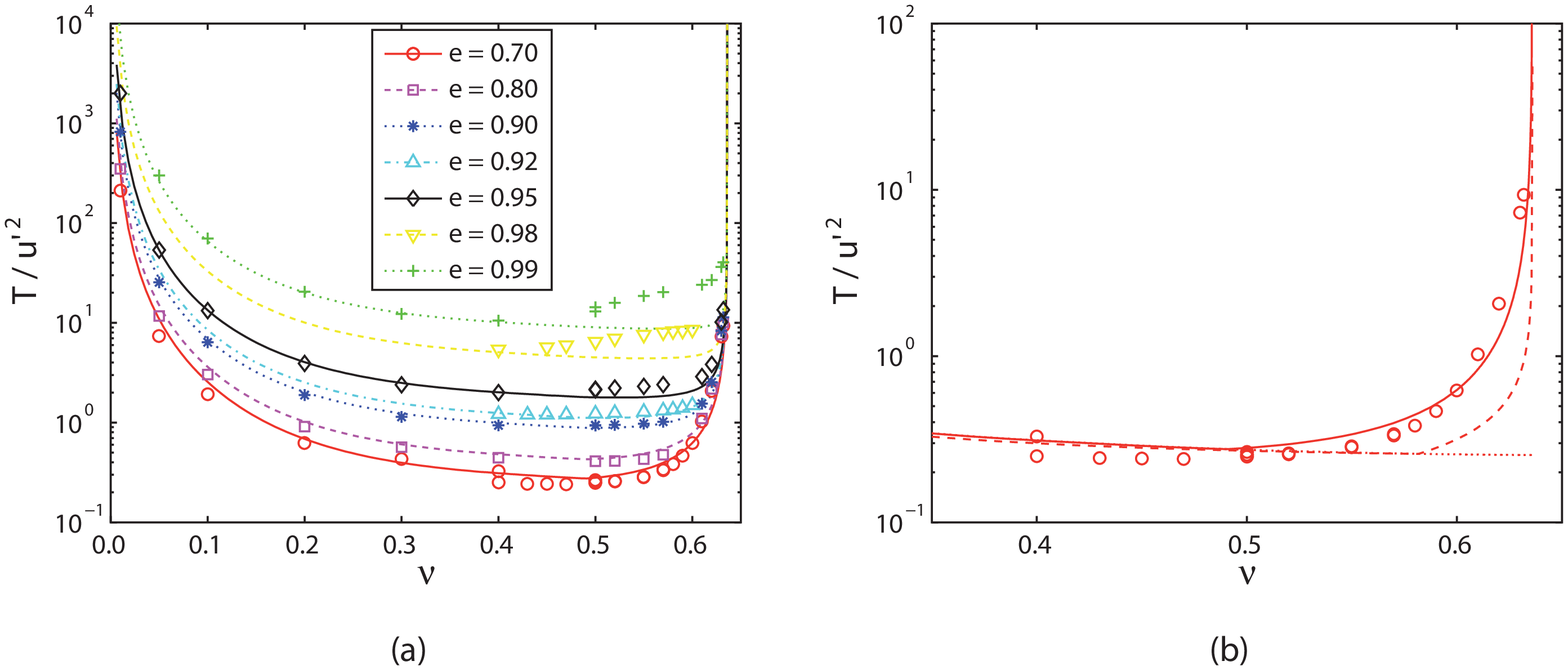}
\caption{(a) Numerical (symbols, after \citet{mit2007} and \citet{chi2013}) and theoretical (lines, \Eq\eqref{ssf_T}) ratio of granular temperature to the square of the shear rate as a function of the volume fraction, for different values of the coefficient of restitution. (b) Same as in \Fig\ref{fig_T}(a) for the case $e=0.7$. The dotted line represents the present theory when $L=1$, while the dashed line the theory of \citet{chi2013}.}
\label{fig_T}
\end{figure}

\begin{figure}[!h]
\centering
\includegraphics[width=0.7\textwidth]{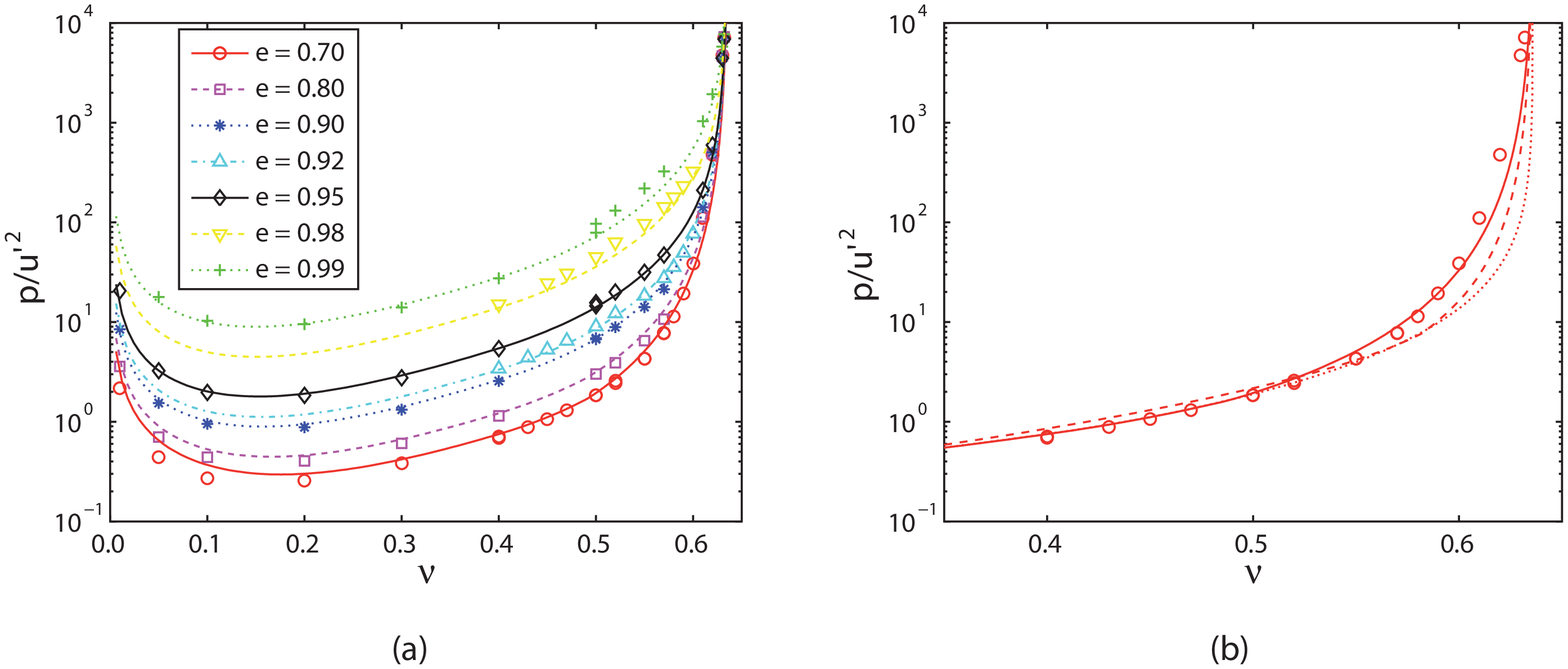}
\caption{(a) Numerical (symbols, after \citet{mit2007} and \citet{chi2013}) and theoretical (lines, \Eq\eqref{ssf_pressure}) ratio of pressure to the square of the shear rate as a function of the volume fraction, for different values of the coefficient of restitution. (b) Same as in \Fig\ref{fig_p}(a) for the case $e=0.7$. The dotted line represents the present theory when $L=1$, while the dashed line the theory of \citet{chi2013}.}
\label{fig_p}
\end{figure}

\begin{figure}[!h]
\centering
\includegraphics[width=0.7\textwidth]{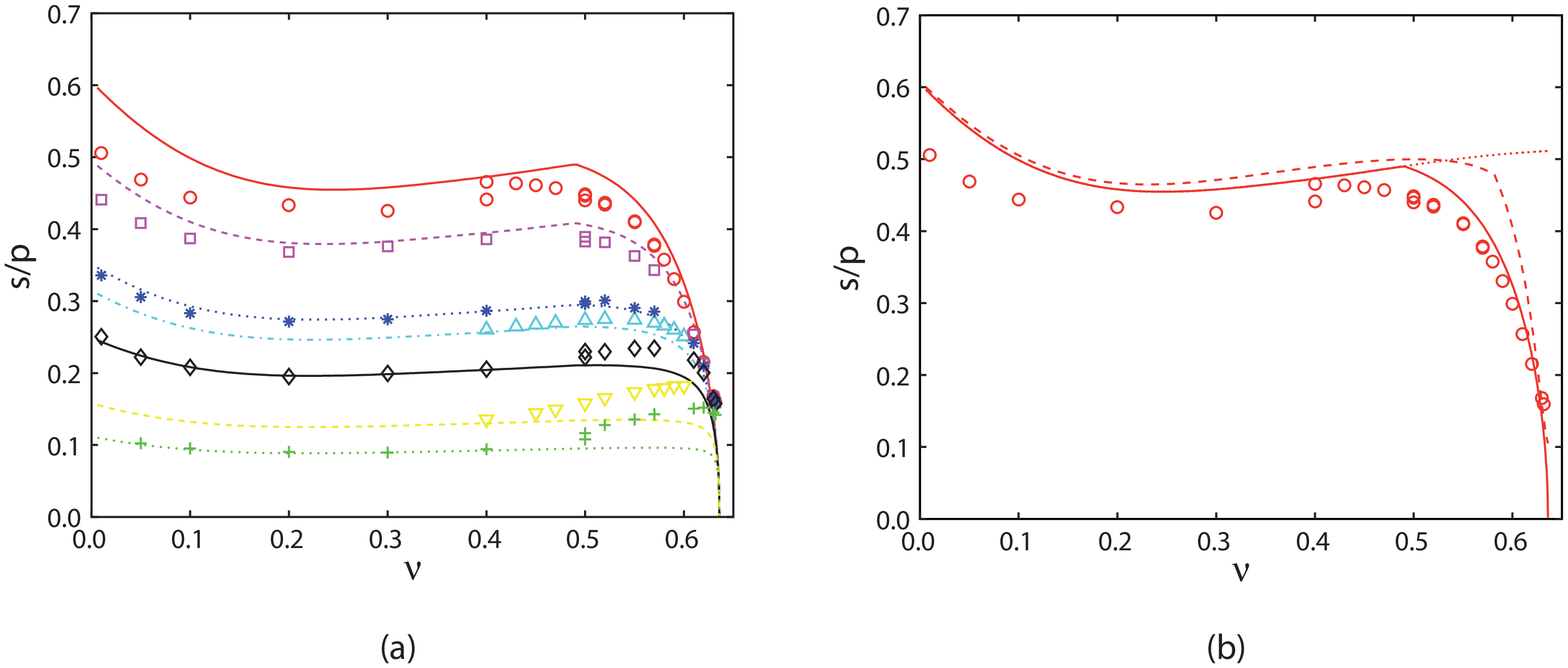}
\caption{(a) Numerical (symbols, after \citet{mit2007} and \citet{chi2013}) and theoretical (lines, \Eq\eqref{ssf_stressratio}) ratio of shear stress to the pressure as a function of the volume fraction, for different values of the coefficient of restitution. (b) Same as in \Fig\ref{fig_stressratio}(a) for the case $e=0.7$. The dotted line represents the present theory when $L=1$, while the dashed line the theory of \citet{chi2013}.}
\label{fig_stressratio}
\end{figure}

\noindent The quantities $T/{u'}^2$, $p/{u'}^2$ and $s/p$, obtained from the numerical simulations of \citet{mit2007} and \citet{chi2013}, are shown in \Figs\ref{fig_T}(a), \ref{fig_p}(a) and \ref{fig_stressratio}(a), respectively, for different values of the coefficient of restitution. The lines represent \Eqs\eqref{ssf_T}, \eqref{ssf_pressure} and \eqref{ssf_stressratio} with the radial distribution function given by \Eq\eqref{g0_theo} and the correlation length given by \Eq\eqref{ssf_L2}. In \Figs\ref{fig_T}(b), \ref{fig_p}(b) and \ref{fig_stressratio}(b) we also plot, for the case $e=0.7$, the predictions of the present theory if the breaking of the molecular chaos is not accounted for (i.e., $L=1$) and the predictions from the theory of \citet{chi2013}.

%...........................................
%...........................................

%...........................................
%...........................................

%...........................................
%...........................................

%...........................................
%...........................................

\noindent Except for large coefficients of restitution ($e > 0.95$), the granular temperature, the pressure and the stress ratio are well predicted by kinetic theory in the entire range of volume fraction, if the expressions \eqref{g0_theo} for $g_0$ and \eqref{ssf_L2} for $L$ are adopted. 
Replacing \Eq\eqref{g0_theo} with \Eq\eqref{torquato} would allow a good fitting also 
for the case of nearly elastic particles ($e > 0.95$).

\begin{figure}[!h]
\centering
\includegraphics[width=0.7\textwidth]{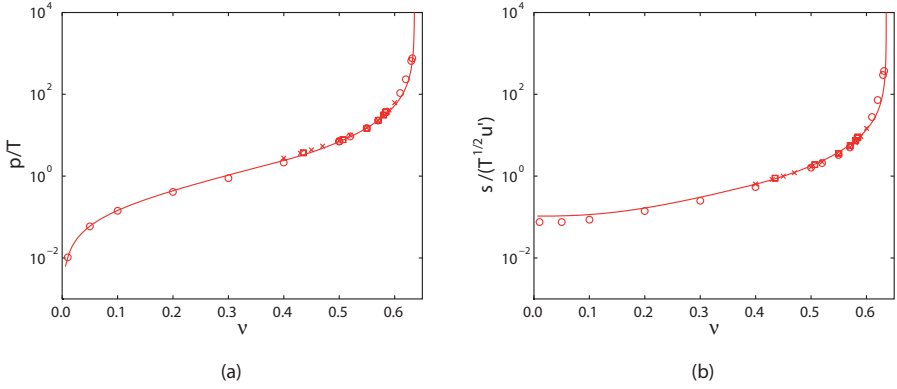}
\caption{Numerical (after \citet{mit2007}, crosses; \citet{chi2013}, circles; present SS-DEM simulations, squares) quantities  $p/T$ (a) and $s/\left( T^{1/2} u' \right)$ (b) as functions of the volume fraction for $e = 0.70$, compared with the theoretical expression of $f_1$ and $f_2$ of \Tab\ref{tab1} (lines).}
\label{fig_f1_f2}
\end{figure}

\noindent Finally, \Figs\ref{fig_f1_f2}(a) and \ref{fig_f1_f2}(b) depict, respectively, the quantity $p/T$ and $s/\left( T^{1/2} u' \right)$ as functions of the volume fraction, where $p$, $T$, $s$ and $u'$ are those measured by \citet{mit2007} and \citet{chi2013}, when $e = 0.70$, together with the theoretical expressions of $f_1$ and $f_2$ of \Tab\ref{tab1}, with $g_0$ given by \Eq\eqref{g0_theo}.   
Also the data obtained from the present SS-DEM simulations on bounded shear granular flows (with $e = 0.70$ and $\psi = \pi/5$) are shown. All the numerical data collapse, independently of the simulation method and the flow configuration, and are in very good agreement with the theoretical curves. Similar agreement is obtained for other values of the coefficient of restitution. In particular, \Fig\ref{fig_f1_f2}(b) indicates that there is no need to modify the constitutive relation of the shear stress of kinetic theory, at least if the particles are sufficiently inelastic. \cite{ber2013} We now compare the results of the numerical integration of \Eqs\eqref{dudy}-\eqref{dmdy}, with the SS-DEM simulations in terms of profiles of volume fraction, velocity and granular temperature, distinguishing between small and large bumpiness.

%...........................................
%...........................................

%.....................................................................................................................
\subsection{Small bumpiness}

\begin{figure}[!h]
\centering
\includegraphics[width=0.7\textwidth]{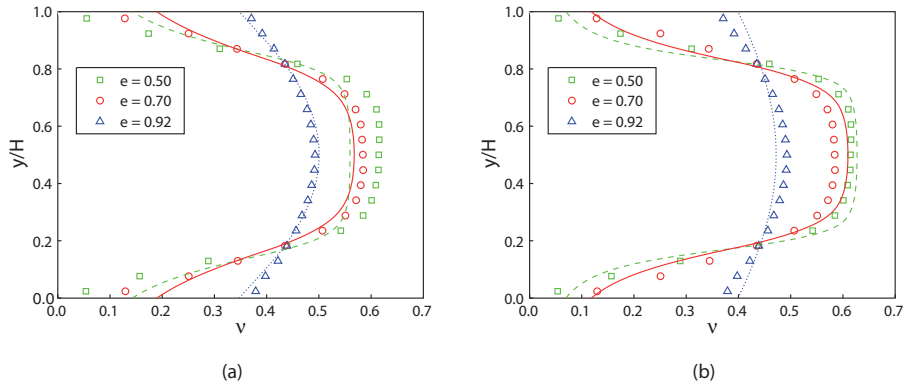}
\caption{Profile of volume fraction obtained from the present SS-DEM simulations (symbols) for $\psi = \pi/5$, $\bar\nu = 0.45$ and various coefficients of restitution. The data are compared with the numerical integration of \Eqs\eqref{dudy}-\eqref{dmdy} for $e = 0.50$ (dashed line), $e = 0.70$ (solid line) and $e = 0.92$ (dot-dashed line) when: (a) the boundary condition on the slip velocity is \Eq\eqref{slipvelocity}; (b) the boundary condition on the slip velocity is \Eq\eqref{h_corr}.}
\label{nu_pi5}
\end{figure}

\begin{figure}[!h]
\centering
\includegraphics[width=0.7\textwidth]{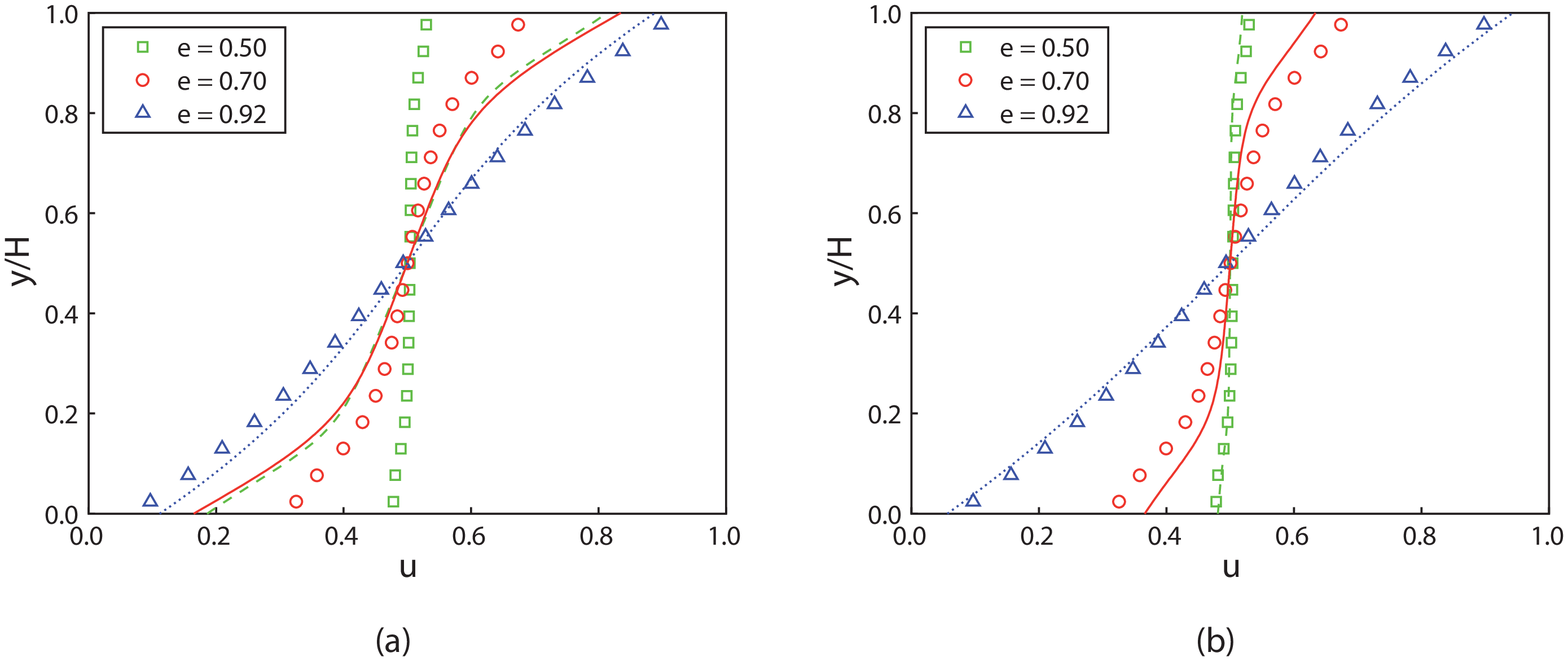}
\caption{Same as in \Fig\ref{nu_pi5}, but for the profile of velocity.}
\label{u_pi5}
\end{figure}

\begin{figure}[!h]
\centering
\includegraphics[width=0.7\textwidth]{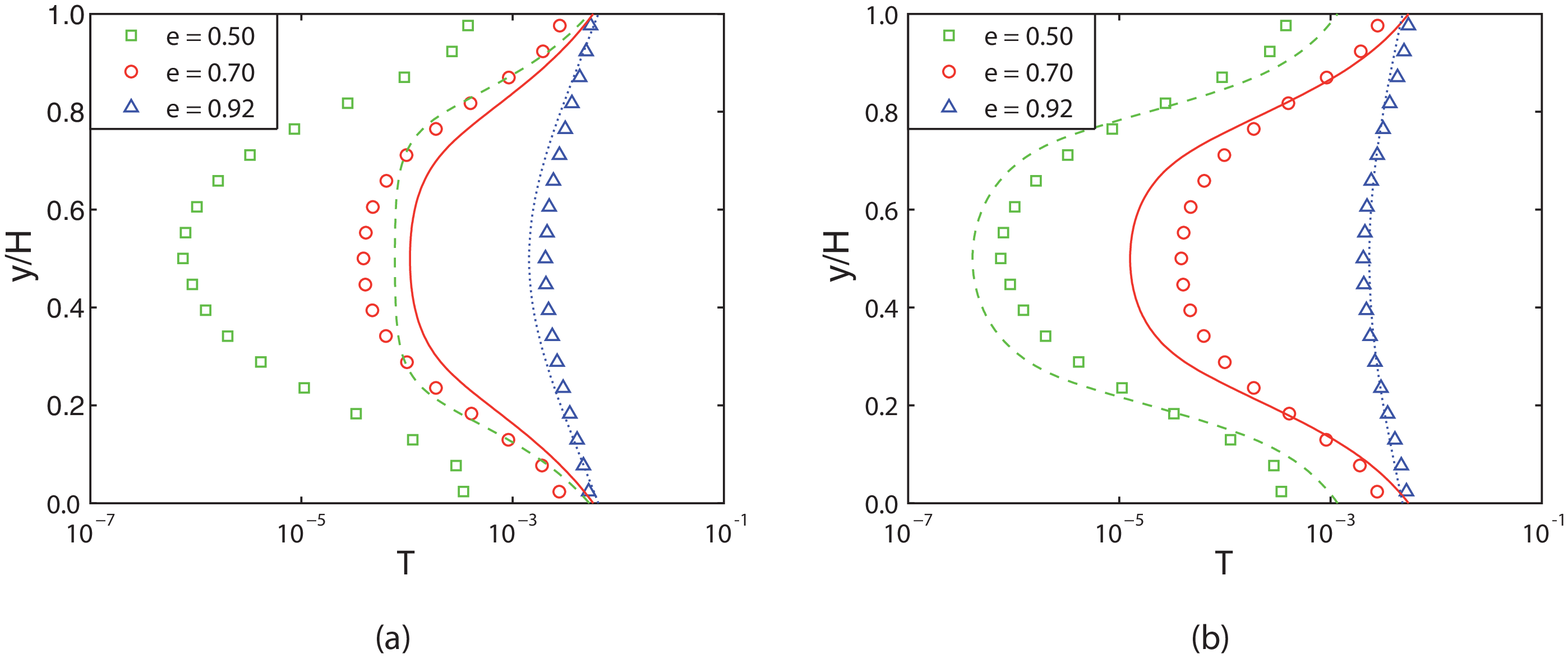}
\caption{Same as in \Fig\ref{nu_pi5}, but for the profile of granular temperature.}
\label{T_pi5}
\end{figure}

\begin{figure}[!h]
\centering
\includegraphics[width=0.7\textwidth]{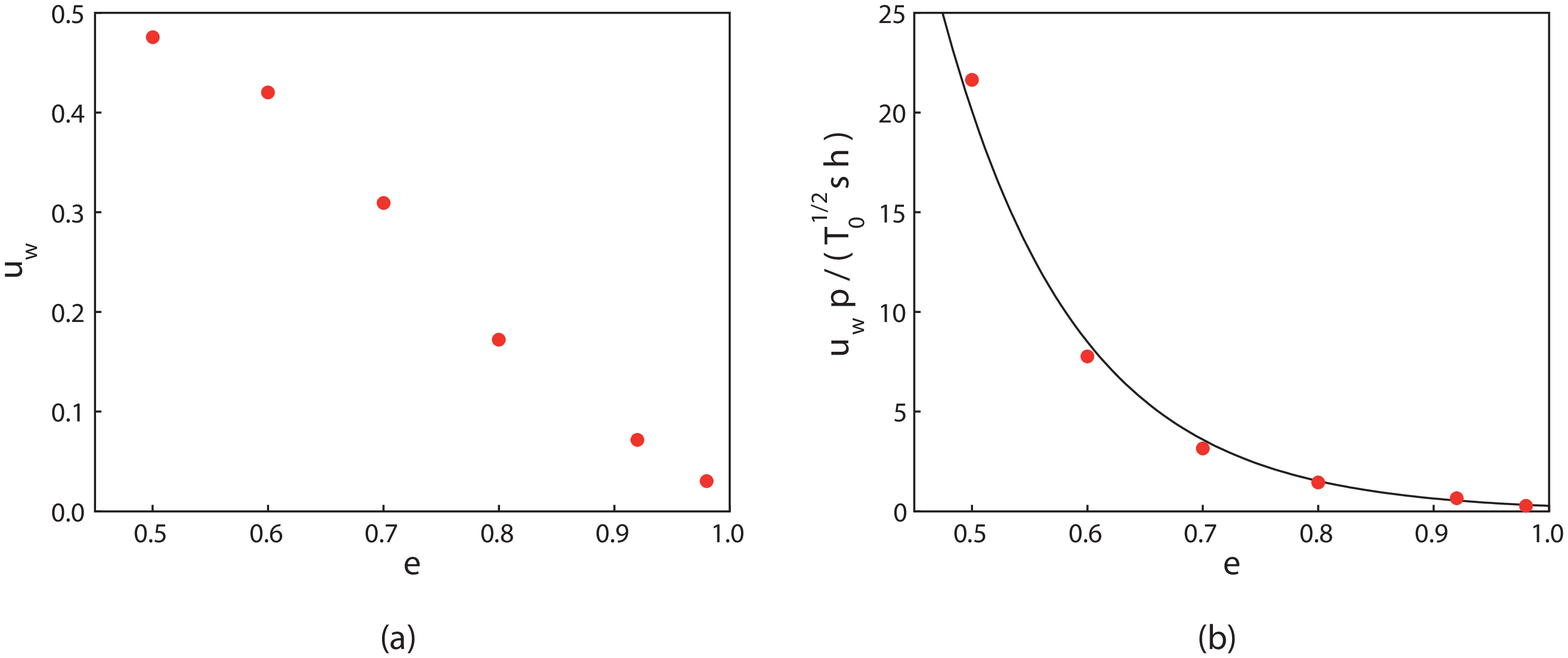}
\caption{(a) Slip velocity as a function of the coefficient of restitution obtained from the present SS-DEM simulation when $\psi=\pi/5$. (b) Correction for the theoretical expression of the coefficient $h$ given in \Eq\eqref{func_h} obtained from the present SS-DEM simulations. The solid line represents \Eq\eqref{h_corr}.}
\label{fig_uw}
\end{figure}

\noindent \Figs\ref{nu_pi5}(a), \ref{u_pi5}(a) and \ref{T_pi5}(a) show that, at small bumpiness ($\psi=\pi/5$), and using the boundary conditions of \citet{ric1988}, EKT only qualitatively reproduces the SS-DEM results, when $\bar\nu=0.45$ as in the simulations. Those boundary conditions were developed for nearly elastic particles. Actually, the slip velocity and the volume fraction are underestimated, and the granular temperature is strongly overestimated when the coefficient of restitution is far from unity. In general, the SS-DEM simulations show that the volume fraction increases with the distance from the wall (\Fig\ref{nu_pi5}a), and the walls are always ``hotter'' than the interior (\Fig\ref{T_pi5}a), i.e., the boundaries are energetic (the fluctuating energy flux is directed towards the interior of the flow); for very inelastic particles, a dense core surrounded by two more dilute layers appear (\Fig\ref{nu_pi5}a). Also, the slip velocity increases as the coefficient of restitution decreases: for $e = 0.50$ the granular material roughly moves as a plug (\Fig\ref{u_pi5}a). \Fig\ref{fig_uw}(a) depicts the value of the slip velocity $u_w$ as a function of the coefficient of restitution. For $e=0.5$, the slip velocity approaches the value 0.5, for which there is a condition of perfect slip at the walls: in that case, the particles do not touch the walls, so that no exchange of energy with the boundaries is possible. This is the reason why, for $e$ lower than 0.5, the energy initially put into the system is entirely dissipated in collisions and the evolution of the mean granular temperature obeys the Haff's law (\Fig\ref{evolution_pi5}). \Fig\ref{fig_uw}(b) shows the ratio of the quantity $u_wp/\left(T_0^{1/2}s\right)$ obtained from the SS-DEM simulations to the coefficient $h$ obtained from \Eq\eqref{func_h} using $\psi=\pi/5$ and the numerical values of the volume fraction at the walls. The boundary condition on the slip velocity of \citet{ric1988} must be corrected in order to reproduce the measurements. On the basis of best fitting, we propose to use
\begin{equation}\label{h_corr}
\frac{u_w}{T_0^{1/2}s/p} = h\exp(7.3-8.6e).
\end{equation}
which represents the solid line in \Fig\ref{fig_uw}(b). If we employ \Eq\eqref{h_corr} instead of \Eq\eqref{slipvelocity} as a boundary condition, when numerically integrating the equations of EKT, the agreement with the numerical simulations is remarkable even in the case of very inelastic particles (\Figs\ref{nu_pi5}b, \ref{u_pi5}b and \ref{T_pi5}b). We expect the numerical coefficients in \Eq\eqref{h_corr} to depend on the bumpiness and, perhaps, the particle stiffness. We postpone to future works a systematic investigation on the role of those quantities in determining the correction to the slip velocity.

%...........................................
%...........................................

%...........................................
%...........................................

%...........................................
%...........................................

%...........................................
%...........................................

%.....................................................................................................................
\subsection{Large bumpiness}

\begin{figure}[!h]
\centering
\includegraphics[width=0.7\textwidth]{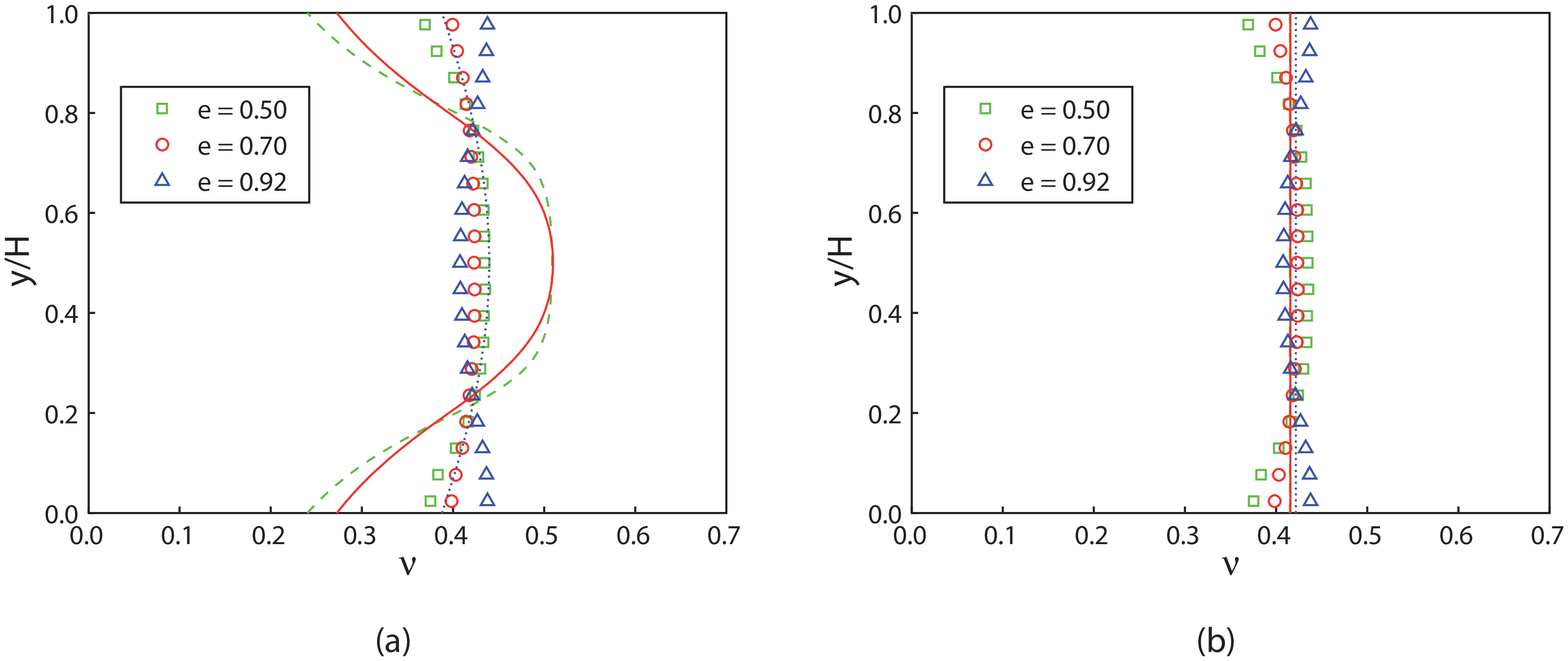}
\caption{Profile of volume fraction obtained from the present SS-DEM simulations (symbols) for $\psi = \pi/3$ and various coefficients of restitution. The data are compared with the numerical integration of \Eqs\eqref{dudy}-\eqref{dmdy} for $e = 0.50$ (dashed line), $e = 0.70$ (solid line) and $e = 0.92$ (dot-dashed line) when: (a) the boundary conditions are \Eq\eqref{slipvelocity} and \Eq\eqref{flux_bc}; (b) the boundary conditions are $u_w=Q_w=0$. In both cases, the mean volume fraction is that measured in the simulations.}
\label{nu_pi3}
\end{figure}

\begin{figure}[!h]
\centering
\includegraphics[width=0.7\textwidth]{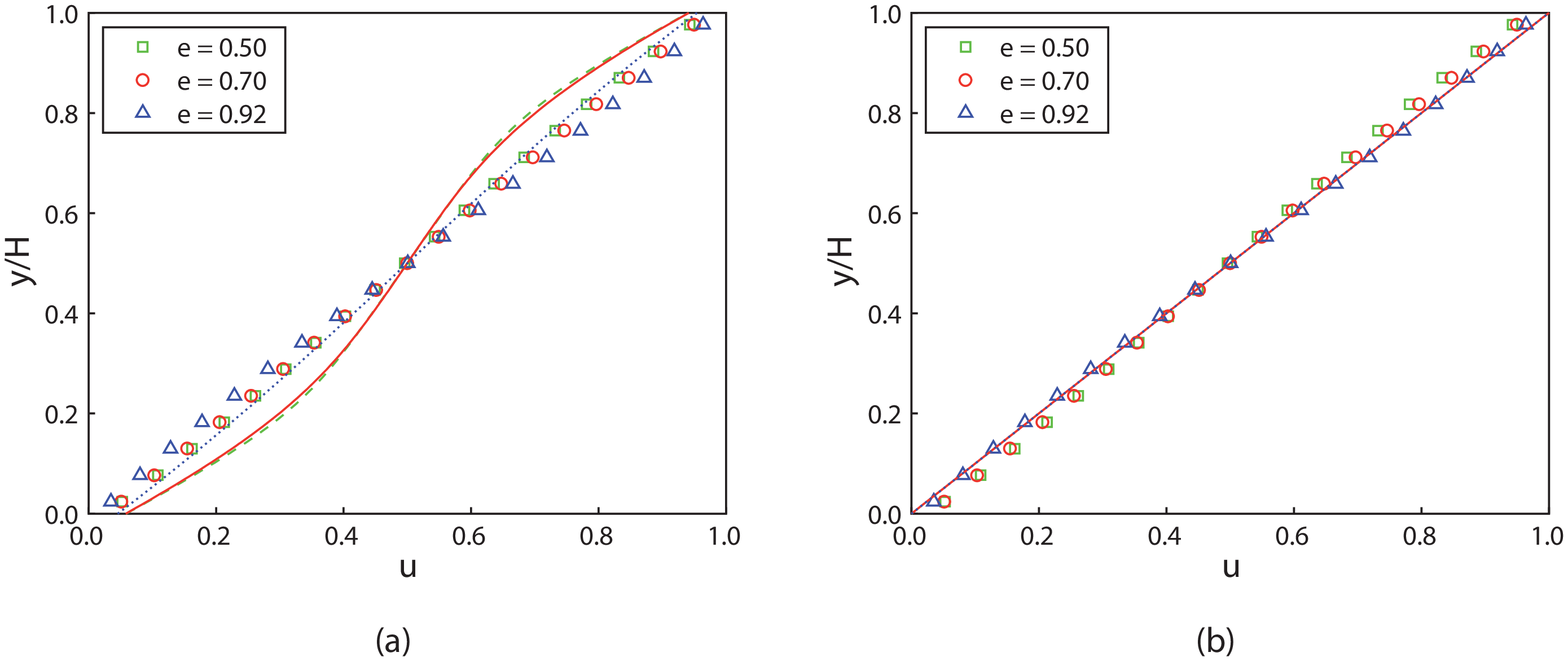}
\caption{Same as in \Fig\ref{nu_pi3}, but for the profile of velocity.}
\label{u_pi3}
\end{figure}

\begin{figure}[!h]
\centering
\includegraphics[width=0.7\textwidth]{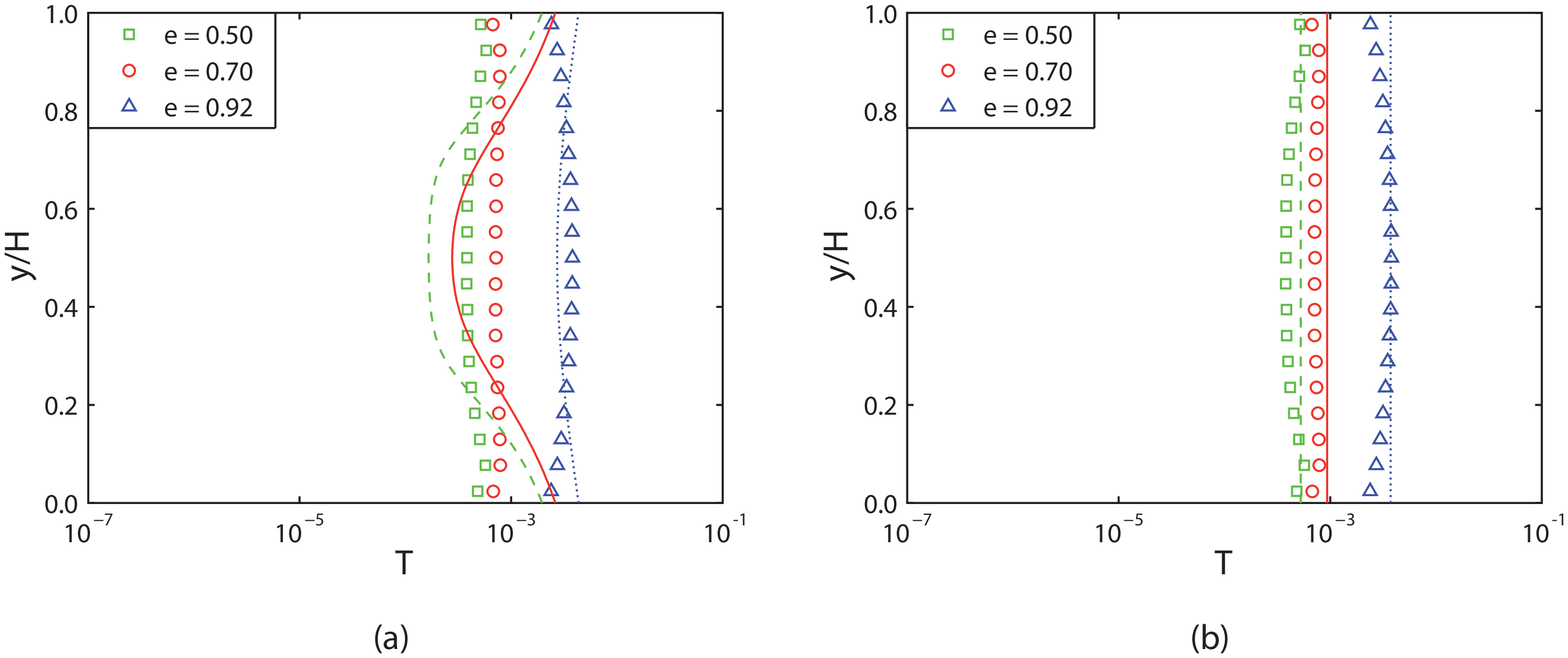}
\caption{Same as in \Fig\ref{nu_pi3}, but for the profile of granular temperature.}
\label{T_pi3}
\end{figure}

\begin{figure}[!h]
\centering
\includegraphics[width=0.7\textwidth]{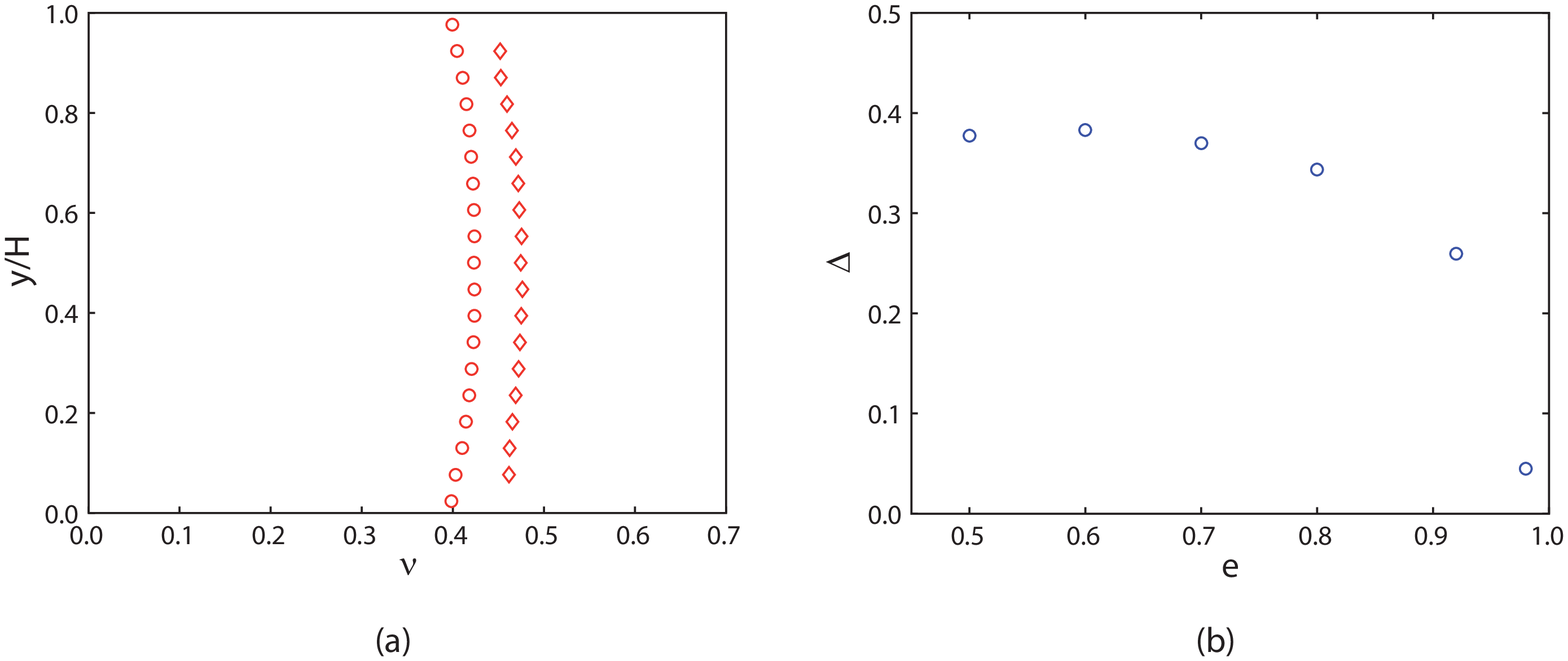}
\caption{(a) Profile of volume fraction obtained from the present SS-DEM simulations with ordered ($\psi = \pi/3$, circles) and disordered (diamonds) bumpy walls, when $e=0.7$ and $N = 3132$. (b) Thickness of the trapped particle layer as a function of the coefficient of restitution when $\psi = \pi/3$.}
\label{pi3_dis}
\end{figure}

\begin{figure}[!h]
\centering
\includegraphics[width=0.7\textwidth]{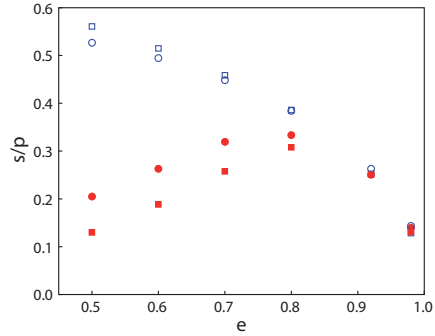}
\caption{Stress ratio $s/p$ as a function of the coefficient of restitution obtained from the SS-DEM simulations when $\psi = \pi/5$ (filled circles) and $\psi = \pi/3$ (open circles), and from the numerical integration of \Eqs\eqref{dudy}-\eqref{dmdy} with the proposed modifications of the boundary conditions ($\psi = \pi/5$, filled squares; and $\psi = \pi/3$, open squares).}
\label{mu_fig}
\end{figure}

\noindent The SS-DEM simulations indicate (\Figs\ref{nu_pi3}, \ref{u_pi3} and \ref{T_pi3}) that, at large bumpiness ($\psi = \pi/3$), the volume fraction and the granular temperature are rather uniform, and the velocity profile is linearly distributed with zero slip velocity (\Fig\ref{u_pi3}), as for simple shear flows. Predictions of EKT in the case $\psi=\pi/3$ when the boundary conditions \Eqs\eqref{slipvelocity} and \eqref{flux_bc} are employed strongly disagree with the SS-DEM results (\Figs\ref{nu_pi3}a, \ref{u_pi3}a and \ref{T_pi3}a). Visual observation of the particle motion suggests that for large enough bumpiness, some of the flowing particles get stuck in the gaps between the particles glued at the walls; those trapped particles contribute then to create a ``disordered'' bumpy wall, similar to that employed in the numerical simulations of \citet{sil2001}, which is far less energetic than the ``ordered'' bumpy wall of \citet{ric1988}. In the cases $e = 0.70$ and 0.92, the walls are even slightly colder than the interior (\Fig\ref{T_pi3}a), i.e., the boundaries are dissipative (the fluctuating energy flux is directed towards the walls). If we use the mean volume fraction obtained by averaging the SS-DEM profiles and $u_w=Q_w=0$ instead of \Eqs\eqref{slipvelocity} and \eqref{flux_bc}, i.e., we assume that the boundaries are neutral (they do not furnish nor subtract fluctuating energy), as boundary conditions, the numerical integration of EKT, which coincides with the analytical solution of simple shear flows, provides a fairly good agreement with the SS-DEM simulations (\Figs\ref{nu_pi3}b, \ref{u_pi3}b and \ref{T_pi3}b). To check our intuition about the particles being trapped at the walls, we have also performed SS-DEM simulations, with $e=0.7$, when random conformations of particles are glued at the walls (the details for the generation of this kind of boundaries are given in \citet{sil2001}). The distribution of the volume fraction (and of the other quantities, not shown here for sake of brevity) is very similar to the case $\psi=\pi/3$ (\Fig\ref{pi3_dis}a). The mean volume fraction is different in the two cases, because the space accessible to the flowing particles, whose number is constant and equal to 3132, is different. Also, the fact that the mean volume fraction measured in the SS-DEM simulatons $\bar{\nu}_{\dem}$ is, in general, less than the theoretical value 0.44, that would characaterize the $\psi=\pi/3$ case when $N=3132$, is an indication of particle trapping. Indeed, a rough estimate of the thickness $\Delta$ of this trapped particle layer is
\begin{equation}\label{Delta}
\Delta=\frac{N}{2L_xL_z}\left(1-\frac{\bar{\nu}_{\dem}}{0.44}\right).
\end{equation}
\Fig\ref{pi3_dis}(b) shows that $\Delta$ goes to zero as $e$ approaches one. Also, the thickness $\Delta$ saturates to a constant value for coefficients of restitution lower than 0.7. Once again, we postpone to future works the generalization of these findings to other values of the bumpiness and the particle stiffness.\\
\noindent Finally, \Fig\ref{mu_fig} shows the influence of the coefficient of restitution on the stress ratio, $s/p$.
Contrary to results reported for 2D plane shear flows of frictional grains submitted to imposed pressure,\cite{dac2005} the coefficient of restitution strongly affects the stress ratio. 
In the range $0.50 \leq e \leq 0.98$, the stress ratio obtained from the present SS-DEM simulations is a decreasing function of the coefficient of restitution for large bumpiness ($\psi = \pi/3$); while $s/p$ has a maximum around $e = 0.80$ for small bumpiness ($\psi = \pi/5$). 
The predictions of EKT when \Eqs\eqref{h_corr} and \eqref{flux_bc} are employed as boundary conditions for $\psi=\pi/5$, and $u_w=Q_w=0$ for $\psi=\pi/3$ are, once again, in a fairly good agreement with the simulations. The drop in the stress ratio for small bumpiness and coefficients of restitution less than 0.8 is due to the already mentioned increasing of the slip velocity, with the corresponding approaching to the Homogeneous Cooling State, in which the shear stress, and consequently the stress ratio, vanishes.

%...........................................
%...........................................

%...........................................
%...........................................

%...........................................
%...........................................

%...........................................
%...........................................

%...........................................
%...........................................

%-----------------------------------------------------------------
%-----------------------------------------------------------------
\section{Conclusions}\label{conclusions}

In this paper, the Extended Kinetic Theory is numerically solved for the shear flows of identical, frictionless particles bounded between two parallel, bumpy planes, at constant volume (plane shear flow). The bumpiness is due to spheres identical to those of the flow, glued at the walls in a regularly spaced, hexagonal array. 
The numerical solutions are compared with 3D SS-DEM simulations, and the roles of the coefficient of restitution and the bumpiness of the boundaries are investigated. 
We propose an expression for the radial distribution function to be used when $e \leq 0.95$ which coincides with the Carnahan and Starling's\cite{car1969} at small volume fraction, and diverges as the volume fraction approaches the shear rigidity as the Torquato's,\cite{tor1995} but, unlike the latter, its derivative is continuous in the entire range of volume fraction. We have shown that the proposed expression fits well the results of ED and SS-DEM simulations of simple shear flows. 
Also, we adopt a recently suggested expression for the correlation length in the dissipation rate of fluctuating energy, which depends only on the coefficient of restitution.
At small bumpiness, the SS-DEM simulations show that the volume fraction increases with the distance from the wall, for every value of the coefficient of restitution, and the wall is always ``hotter'' than the interior. The slip velocity at the boundaries decreases with the elasticity of the particles, and, for coefficients of restitution less than 0.5, the slip is perfect: the boundaries do not touch the flowing particles, so that the system evolves accordingly to the Homogeneous Cooling State (it is not possible to obtain a steady shear flow). Also, the measured stress ratio is a non monotonic function of the coefficient of restitution, and reaches a maximum for $e = 0.80$. 
The results of the numerical integration of EKT agree well with the simulations, if a correction to the expression of the slip velocity depending on the coefficient of restitution is introduced in the boundary conditions derived for nearly elastic particles by \citet{ric1988}. 
At large bumpiness, the SS-DEM simulations show nearly uniform profiles of volume fraction and granular temperature, and linear distributions of the velocity field, as for simple shear flows. 
This is due to the fact that, when the gaps between the spheres glued at the walls are large enough, some of the flowing particles get stuck, making the bumpy wall more ``disordered'', and, then, more dissipative than expected. Even in the case of large bumpiness, EKT is able to reproduce the simulation results, if both the slip velocity and the fluctuating energy flux at the walls are taken to be zero.\\
Summarizing, we have shown that Extended Kinetic Theory has the capability of quantitatively reproducing the flow of frictionless spheres in the entire range of volume fraction for which the collisions can be considered nearly instantaneous and random (i.e., the entire fluid-like regime of granular flows). Tests of proposed extensions to EKT to deal with friction, non-instantaneous collisions and enduring contacts will be the subject of future works.

%-----------------------------------------------------------------
%-----------------------------------------------------------------
% bibliography
\bibliography{biblio}

\end{document}